\documentclass[editedvolume,numreferences]{crckbked}

\usepackage{graphicx}

\def \be{\begin{equation}}
\def \ee{\end{equation}}
\def \bmlett{\begin{mathletters}}
\def \emlett{\end{mathletters}}
\def \r{{\bf r}}

\def \ve{\varepsilon}

\def \NN{{\mathcal N}}

\def \ua{\uparrow}
\def \da{\downarrow}
\def \ra{\rightarrow}

\def\kbt{k_{\rm B}T}
\def\be{\begin{equation}}
\def\ee{\end{equation}}
\def\ketn{|n\rangle}
\def\bran{\langle n|}

\def\ketg{|g\rangle}
\def\kete{|e\rangle}

\def\brag{\langle g|}
\def\gammaup{\Gamma_\uparrow}
\def\gammadown{\Gamma_\downarrow}
\def\w01{\omega_{01}}
\def\r0{R_0}

\begin{document}
\begin{opening}
\title{Qubits as Spectrometers of Quantum Noise}
\author{R.J. Schoelkopf}
\author{A.A. Clerk}
\author{S.M. Girvin}
\author{K.W. Lehnert}
\author{M.H. Devoret}
\institute{Departments of Applied Physics and
Physics\\Yale University, PO Box 208284, New Haven, CT 06520-8284}
\end{opening}

\section{Introduction}

Electrical engineers and physicists are naturally very interested
in the noise of circuits, amplifiers and detectors.  This noise
has many origins, some of which are completely unavoidable.  For
example, a dissipative element (a resistor) at finite temperature
inevitably generates Johnson noise. Engineers long ago developed
spectrum analyzers to measure the intensity of this noise. Roughly
speaking, these spectrum analyzers consist of a resonant circuit
to select a particular frequency of interest, followed by an
amplifier and square law detector (e.g. a diode rectifier) which
measures the mean square amplitude of the signal at that
frequency.

With the advent of very high frequency electronics operating at
low temperatures, we have entered a new regime $\hbar\omega >
\kbt$, where quantum mechanics plays an important role and one has
to begin to think about {\em quantum noise} and quantum-limited
amplifiers and detectors.  This topic is well-studied in the
quantum optics community and is also commonplace in the radio
astronomy community.  It has recently become of importance in
connection with quantum computation and the construction of
mesoscopic electrical circuits which act like artificial atoms
with quantized energy levels.  It is also important for
understanding the quantum measurement process in mesoscopic
systems.

In a classical picture, the intensity of Johnson noise from a
resistor vanishes linearly with temperature because thermal
fluctuations of the charge carriers cease at zero temperature. One
knows from quantum mechanics, however, that there are quantum
fluctuations even at zero temperature, due to zero-point motion.
Zero-point motion is a notion from quantum mechanics that is
frequently misunderstood. One might wonder, for example, whether
it is physically possible to use a spectrum analyzer to detect the
zero-point motion. The answer is quite definitely yes, if we use a
quantum system! Consider for example a hydrogen atom in the 2p
excited state lying 3/4 of a Rydberg above the 1s ground state. We
know that this state is unstable and has a lifetime of only about
1 ns before it decays to the ground state and emits an ultraviolet
photon.  This spontaneous decay is a natural consequence of the
zero-point motion of the electromagnetic fields in the vacuum
surrounding the atom.  In fact, the rate of spontaneous decay
gives a simple way in which to {\em measure} this zero point
motion of the vacuum. Placing the atom in a resonant cavity can
modify the strength of the noise at the transition frequency, and
this effect can be measured via a change in the decay rate.

At finite temperature, the vacuum will contain blackbody photons
which will increase the rate of decay due to stimulated emission
and also cause transitions in the reverse direction,
1s~$\rightarrow$~2p, by photon absorption.  With these ideas in
mind, it is now possible to see how to build a quantum spectrum
analyzer.

The remainder of this article is organized as follows. First we
describe the general concept of a two-level system as a quantum
spectrum analyzer. We next review the Caldeira-Leggett formalism
for the modelling of a dissipative circuit element, such as a
resistor, and its associated quantum noise. Then, a brief
discussion of the single Cooper-pair box, a circuit which behaves
as a two-level system or qubit, is given. We then discuss the
effects of a dissipative electromagnetic environment on the box,
and treat the case of a simple linear, but {\em nonequilibrium}
environment, consisting of a classical tunnel junction which
produces shot noise under bias. Finally, we describe a theoretical
technique for calculating the properties of a Cooper-pair box
coupled to a measurement system, which will be a {\em nonlinear,
nonequilibrium} device, such as a single-electron transistor.
Equivalently, this allows one to calculate the full {\em quantum
noise spectrum} of the measurement device. Results of this
calculation for the case of a normal SET are presented.

\section{Two-level systems as spectrum analyzers \label{TLSanalyzer}}

Consider a quantum system (atom or electrical circuit) which has
its two lowest energy levels $\epsilon _{0}$ and $\epsilon _{1}$
separated by energy $E_{01}=\hbar \w01$. We suppose for simplicity
that all the other levels are far away in energy and can be
ignored. The states of any two-level system can be mapped onto the
states of a fictitious spin-1/2 particle since such a spin also
has only two states in its Hilbert space. With spin up
representing the ground state ($|g\rangle $) and spin down
representing the excited state~($|e\rangle $), the Hamiltonian is
(taking the zero of energy to be the center of gravity of the two
levels)
\begin{equation}
H_{0}=-\frac{\hbar \w01}{2}\sigma _{z}.
\end{equation}%
In keeping with the discussion above, our goal is to see how the
rate of `spin-flip' transitions induced by an external noise
source can be used to analyze the spectrum of that noise. Suppose
for example that there is a noise source with amplitude $f(t)$
which can cause transitions via the
perturbation\footnote{%
The most general perturbation would also couple to $\sigma _{y}$
but we assume that (as is often, though not always, the case) a
spin coordinate system can be chosen so that the perturbation only
couples to $\sigma _{x}$. Noise coupled to $\sigma _{z}$ commutes
with the Hamiltonian but is nevertheless important in dephasing
coherent superpositions of the two states. We will not discuss
such processes here.}
\begin{equation}
V=Af(t)\sigma _{x},
\end{equation}%
where $A$ is a coupling constant. The variable $f(t)$ represents
the noise source. We can temporarily pretend that $f$ is a
classical variable, although its quantum operator properties will
be forced upon us very soon. For now, only our two-level spectrum
analyzer will be treated quantum mechanically.

We assume that the coupling $A$ is under our control and can be
made small enough that the noise can be treated in lowest order
perturbation theory. We take the state of the two-level system to
be
\begin{equation}
|\psi (t)\rangle =\left(
\begin{array}{c}
\alpha _{g}(t) \\
\alpha _{e}(t)%
\end{array}%
\right) .
\end{equation}%
In the interaction representation, first-order time-dependent
perturbation theory gives
\begin{equation}
|\psi _{\mathrm{I}}(t)\rangle =|\psi (0)\rangle -\frac{i}{\hbar }%
\int_{0}^{t}d\tau \,\,\hat{V}(\tau )|\psi (0)\rangle .
\end{equation}%
If we initially prepare the two-level system in its ground state
then the amplitude to find it in the excited state at time $t$ is
\begin{eqnarray}
\alpha _{e} &=&-\frac{iA}{\hbar }\int_{0}^{t}d\tau \,\,\langle e |
\hat{\sigma}_{x}(\tau)|g \rangle f(\tau )+O(A^{2}), \\
&=&-\frac{iA}{\hbar }\int_{0}^{t}d\tau \,\,e^{i\omega _{01}\tau
}f(\tau )+O(A^{2}).
\end{eqnarray}
We can now compute the probability
\begin{equation}
p_{e}(t)\equiv |\alpha _{e}|^{2}=\frac{A^{2}}{\hbar ^{2}}\int_{0}^{t}%
\int_{0}^{t}d\tau _{1}d\tau _{2}\,e^{-i\omega _{01}\left( \tau
_{1}-\tau _{2}\right) }f(\tau _{1})f(\tau _{2})+O(A^{3})
\end{equation} We are actually only interested
on the average time evolution of the system
\begin{equation}
\bar{p}_{e}(t)=\frac{A^{2}}{\hbar
^{2}}\int_{0}^{t}\int_{0}^{t}d\tau _{1}d\tau _{2}\,e^{-i\omega
_{01}\left( \tau _{1}-\tau _{2}\right) }\left\langle f(\tau
_{1})f(\tau _{2})\right\rangle +O(A^{3})
\label{pexciteoft}\end{equation} We can now perform a change of
variables in the integrals, $\tau =\tau _{1}-\tau _{2}$ and $T
=\left( \tau _{1}+\tau _{2}\right) /2$, and we get
\begin{equation}
\bar{p}_{e}(t)=\frac{A^{2}}{\hbar
^{2}}\int_{0}^{t}dT\int_{-B\left( T\right) }^{B\left( T\right)
}d\tau \,e^{-i\omega _{01}\tau }\left\langle f(T+\tau /2)f(T-\tau
/2)\right\rangle +O(A^{3})
\end{equation}
where
\begin{eqnarray*}
B\left( T\right) &=&T~\mathrm{if~}T<t/2 \\
&=&t-T~\mathrm{if~}T>t/2.
\end{eqnarray*}
Let us now suppose that the noise correlation function is
stationary (time translation invariant) and has a finite but small
autocorrelation time $\tau _{f}$. Then for $t\gg \tau _{f}$ we can
set the bound $B\left( T\right) $ to infinity in the last integral
and write
\begin{equation}
\bar{p}_{e}(t)=\frac{A^{2}}{\hbar ^{2}}\int_{0}^{t}dT\int_{-\infty
}^{\infty }d\tau \,e^{-i\omega _{01}\tau }\left\langle f(\tau
)f(0)\right\rangle +O(A^{3})
\end{equation}
The integral over $\tau $ is effectively a sum of a very large
number $N\sim t/\tau _{f}$ of random terms \footnote{The size of
these random terms depends on the variance of $f$ and on the value
of $\w01 \tau_{f}$ For $\w01 \tau_{f}\gg 1$ the size will be
strongly reduced by the rapid phase oscillations of the
exponential in the integrand.} and hence the value undergoes a
random walk as a function of time. Introducing the noise spectral
density
\begin{equation}
S_{f}(\omega )=\int_{-\infty }^{+\infty }d\tau \,e^{i\omega \tau
}\langle f(\tau )f(0)\rangle ,
\end{equation}
we find that the probability to be in the excited state increases \emph{%
linearly} with time,\footnote{Note that for very long times, where
there is a significant depletion of the probability of being in
the initial state, first-order perturbation theory becomes
invalid. However, for sufficiently small $A$, there is a wide
range of times $\tau_f\ll t\ll 1/\Gamma$ for which
Eq.~\ref{lineart} is valid. Eqs.~\ref{gamup} and \ref{gamdown}
then yield well-defined rates which can be used in a master
equation to describe the full dynamics including long times.}
\begin{equation}
\bar{p}_{e}(t)=t\frac{A^{2}}{\hbar ^{2}}S_{f}(-\omega _{01})
\label{lineart}
\end{equation}
The time derivative of the probability gives the transition rate
\begin{equation}
\Gamma _{\uparrow }=\frac{A^{2}}{\hbar ^{2}}S_{f}(-\omega _{01})
\label{gamup}
\end{equation}%
Note that we are taking in this last expression the spectral
density on the negative frequency side. If $f$ were a strictly
classical source $\langle f(\tau )f(0)\rangle $ would be real and
$S_{f}(-\omega _{01})=S_{f}(+\omega _{01})$. However, because as
we discuss below $f$ is actually an operator acting on the
environmental degrees of freedom, $\left[ f(\tau ),f(0)\right] \neq 0$ and $%
S_{f}(-\omega _{01})\neq S_{f}(+\omega _{01})$.

Another possible experiment is to prepare the two-level system in
its excited state and look at the rate of decay into the ground
state. The algebra is identical to that above except that the sign
of the frequency is reversed:
\begin{equation}
\Gamma _{\downarrow }=\frac{A^{2}}{\hbar ^{2}}S_{f}(+\omega
_{01}). \label{gamdown}
\end{equation}
We now see that our two-level system does indeed act as a quantum
spectrum analyzer for the noise. Operationally, we prepare the
system either in its ground state or in its excited state, weakly
couple it to the noise source, and after an appropriate interval
of time (satisfying the above inequalities) simply measure whether
the system is now in its excited state or ground state. Repeating
this protocol over and over again, we can find the probability of
making a transition, and thereby infer the rate and hence the
noise spectral density at positive and negative frequencies. Note
that in contrast with a classical spectrum analyzer, we can
separate the noise spectral density at positive and negative
frequencies from each other since we can separately measure the
downward and upward transition rates. Negative frequency noise
transfers energy \emph{from the noise source to the spectrometer}.
That is, it represents energy emitted by the noise source.
Positive frequency noise transfers energy \emph{from the
spectrometer to the
noise source}.\footnote{%
Unfortunately, there are several conventions in existence for
describing the noise spectral density. It is common in engineering
contexts to use the phrase `spectral density' to mean
$S_{f}(+\omega )+S_{f}(-\omega )$. This is convenient in classical
problems where the two are equal. In quantum contexts, one
sometimes sees the asymmetric part of the noise $S_{f}(+\omega
)-S_{f}(-\omega )$ referred to as the `quantum noise.' We feel it
is simpler and clearer to simply discuss the spectral density for
positive and negative frequencies \emph{separately}, since they
have simple physical interpretations and directly relate to
measurable quantities. This convention is especially useful in
non-equilibrium situations where there is no simple relation
between the spectral densities at positive and negative
frequencies.} In order to exhibit frequency resolution,
$\Delta\omega$, adequate to distinguish these two cases, it is
crucial that the two-level quantum spectrometer have sufficient
phase coherence so that the linewidth of the transitions satisfies
the condition $\w01/\Delta\omega \geq \max[\kbt/\hbar\w01,1]$.

In thermodynamic equilibrium, the transition rates must obey
detailed balance $\Gamma _{\downarrow }/\Gamma _{\uparrow
}=e^{\beta \hbar \w01 }$ in order to give the correct equilibrium
occupancies of the two states of the spectrometer. This implies
that the spectral densities obey
\begin{equation}
S_{f}(+\w01 )=e^{\beta \hbar \w01 }S_{f}(-\w01).
\end{equation}
Without the crucial distinction between positive and negative
frequencies, and the resulting difference in rates, one always
finds that our two level systems are completely unpolarized. If,
however, the noise source is an amplifier or detector biased to be
out of equilibrium, no general relation holds.

We now rigorously treat the quantity $f(\tau )$ as quantum
operator in the Hilbert space of the noise source. The previous
derivation is unchanged, and Eqs.~(\ref{gamup},\ref{gamdown}) are
still valid provided that we interpret the angular brackets in
Eq.~(\ref{pexciteoft}) as representing the quantum statistical
expectation value for the operator correlation (in the absence of
the coupling to the spectrometer)
\begin{equation}
S_{f}(\omega )=\int_{-\infty }^{+\infty }d\tau \,e^{i\omega \tau
}\sum_{\alpha ,\gamma }\rho _{\alpha \alpha }\,\langle \alpha
|f(\tau )|\gamma \rangle \langle \gamma |f(0)|\alpha \rangle
\end{equation}%
where for simplicity we have assumed that (in the absence of the
coupling to the spectrometer) the density matrix is diagonal in
the energy eigenbasis and time-independent (but not necessarily
given by the equilibrium expression). This yields the standard
quantum mechanical expression for the spectral density
\begin{eqnarray}
S_{f}(\omega ) &=&\int_{-\infty }^{+\infty }d\tau \,e^{i\omega
\tau }\sum_{\alpha ,\gamma }\rho _{\alpha \alpha
}\,e^{\frac{i}{\hbar }(\epsilon
_{\alpha }-\epsilon _{\gamma })t}\,|\langle \alpha |f|\gamma \rangle |^{2} \\
&=&2\pi \hbar \sum_{\alpha ,\gamma }\rho _{\alpha \alpha
}\,|\langle \alpha |f|\gamma \rangle |^{2}\delta (\epsilon
_{\gamma }-\epsilon _{\alpha }-\hbar \omega ).
\end{eqnarray}%
Substitution of this into Eqs.~(\ref{gamup},\ref{gamdown}) we
derive the familiar Fermi Golden Rule expressions for the two
transition rates.

In standard courses, one is not normally taught that the
transition rate of a discrete state into a continuum as described
by Fermi's Golden Rule can (and indeed should!) be viewed as
resulting from the continuum acting as a quantum noise source. The
above derivation hopefully provides a motivation for this
interpretation.

One standard model for the continuum is an infinite collection of
harmonic oscillators. The electromagnetic continuum in the
hydrogen atom case mentioned above is a prototypical example. The
vacuum electric field noise coupling to the hydrogen atom has an
extremely short autocorrelation time because the range of mode
frequencies $\omega_\alpha$ (over which the dipole matrix element
coupling the atom to the mode electric field $\vec E_\alpha$ is
significant) is extremely large, ranging from many times smaller
than the transition frequency to many times larger. Thus the
autocorrelation time of the vacuum electric field noise is
considerably less than $10^{- 15}$s, whereas the decay time of the
hydrogen 2p state is about $10^{- 9}$s. Hence the inequalities
needed for the validity of our expressions are very easily
satisfied.

Of course in the final expression for the transition rate, energy
conservation means that only the spectral density at the
transition frequency enters. However, in order for the expression
to be valid (and in order for the transition rate to be time
independent), it is essential that there be a wide range of
available photon frequencies so that the vacuum noise has an
autocorrelation time much shorter than the inverse of the
transition rate.

\section{Quantum Noise from a Resistor\label{SecQuantumResistor}}
Instead of an atom in free space, we might consider a quantum bit
capacitively coupled to a transmission line.  The transmission
line is characterized by an inductance per unit length $\ell$ and
capacitance per unit length $c$.  A semi-infinite transmission
line presents a frequency-independent
 impedance
$ Z = \r0 = \sqrt{{\ell}/{c}} $ at its end and hence acts like an
ideal resistor.  The dissipation is caused by the fact that
currents injected at one end launch waves which travel off to
infinity and do not return.  Very conveniently, however, the
system is simply a large collection of harmonic oscillators (the
normal modes) and hence can be readily quantized.  This
representation of a physical resistor is essentially the one used
by Caldeira and Leggett \cite{caldeiraleggett} in their seminal
studies of the effects of dissipation on tunneling.  The only
difference between this model and the vacuum fluctuations in free
space discussed above is that the relativistic bosons travel in
one dimension and do not carry a polarization label.  This changes
the density of states as a function of frequency, but has no other
essential effect.

The Lagrangian for the system is \be {\cal L} = \int_0^\infty dx\,
\frac{\ell}{2}j^2 - \frac{1}{2c}q^2, \ee where $j(x,t)$ is the
local current density and $q(x,t)$ is the local charge density.
Charge conservation connects these two quantities via the
constraint \be
\partial_x j(x,t) + \partial_t q(x,t) = 0.
\ee We can solve this constraint by defining a new variable \be
\theta(x,t) \equiv \int_0^x dx'\, q(x',t) \label{theta} \ee in
terms of which the current density is $ j(x,t) = -\partial_t
\theta(x,t) $ and the charge density is $ q(x,t) = \partial_x
\theta(x,t)$.  For any well-behaved function $\theta(x,t)$, the
continuity equation is automatically satisfied so there are no
dynamical constraints on the $\theta$ field. In terms of this
field the Lagrangian becomes \be {\cal L} = \int_0^\infty dx\,
\frac{\ell}{2}\left(\partial_t\theta\right)^2 -
\frac{1}{2c}\left(\partial_x\theta\right)^2 \ee The Euler-Lagrange
equation for this Lagrangian is simply the wave equation $ v^2
\partial_x^2 \theta - \partial_t^2\theta = 0 $ where the mode
velocity is $v=1/{\sqrt{\ell c}}$.

>From Eq.~(\ref{theta}) we can deduce that the proper boundary
conditions (in the absence of any coupling to the qubit) for the
$\theta$ field are $\theta(0,t) = \theta(L,t)=0$.  (We have
temporarily made the transmission line have a finite length $L$.)
The normal mode expansion that satisfies these boundary conditions
is \be \theta(x,t) = \sqrt{\frac{2}{L}}\sum_{n=1}^\infty
\varphi_n(t) \sin{\frac{k_n\pi x}{L}}, \ee where $\varphi_n$ is
the normal coordinate and $k_n \equiv \frac{\pi n}{L}$.
Substitution of this form into the Lagrangian and carrying out the
spatial integration yields a set of independent harmonic
oscillators representing the normal modes. \be {\cal L} =
\sum_{n=1}^\infty \frac{\ell}{2} (\dot\varphi_n)^2 - \frac{1}{2c}
k_n^2 \varphi_n^2. \ee From this we can find the momentum $p_n$
canonically conjugate to $\varphi_n$ and quantize the system to
obtain an expression for the voltage at the end of the
transmission line in terms of the mode creation and destruction
operators \be V =
\sqrt{\frac{2}{L}}\frac{1}{c}\partial_x\theta(0,t) =
\frac{1}{c}\sum_{n=1}^\infty k_n
\sqrt{\frac{\hbar}{2\ell\Omega_n}} ( a^\dagger_n + a_n). \ee The
spectral density of voltage fluctuations is then found to be \be
S_{V}(\omega) = 2\pi \frac{2}{L} \sum_{n=1}^\infty
\frac{\hbar\Omega_n}{2c} \big\{ n_\gamma
(\hbar\Omega_n)\delta(\omega+\Omega_n) +
[n_\gamma(\hbar\Omega_n)+1] \delta(\omega-\Omega_n) \big\}, \ee
where $n_\gamma(\hbar\omega)$ is the Bose occupancy factor for a
photon with energy $\hbar\omega$. Taking the limit
$L\rightarrow\infty$ and converting the summation to an integral
yields \be S_{V}(\omega) = 2\r0\hbar|\omega| \big\{
n_\gamma(\hbar|\omega|) \Theta(-\omega) +
[n_\gamma(\hbar\omega)+1] \Theta(\omega) \big\}, \ee where
$\Theta$ is the step function. We see immediately that at zero
temperature there is no noise at negative frequencies because
energy can not be extracted from zero-point motion.  However there
remains noise at positive frequencies indicating that the vacuum
is capable of absorbing energy from the qubit.

A more compact expression for this `two-sided' spectral density of
a resistor is \be S_{V}(\omega) = \frac{2\r0\hbar\omega}
{1-e^{-{\hbar \omega / \kbt}} } ,\ee which reduces to the more
familiar expressions in various limits. For example, in the
classical limit $k_{\rm B}T \gg \hbar\omega$ the spectral density
is equal to the Johnson noise result\footnote{Note again that in
the engineering convention this would be $S_{V}(\omega) =
4\r0k_{\rm B}T$.} \be S_{V}(\omega) = 2\r0k_{\rm B}T, \ee which is
frequency independent, and in the quantum limit it reduces to \be
S_{V}(\omega) = 2\r0\hbar\omega \Theta(\omega). \ee Again, the
step function tells us that the resistor can only absorb energy,
not emit it, at zero temperature.

If we use the engineering convention and add the noise at positive
and negative frequencies we obtain \be S_{V}(\omega) +
S_{V}(-\omega) = 2\r0\hbar\omega \coth\frac{\hbar\omega}{2k_{\rm
B}T} \label{ResistorDensitySymm}\ee for the symmetric part of the
noise, which appears in the quantum fluctuation-dissipation
theorem\cite{CallenWelton}. The antisymmetric part of the noise is
simply \be S_{V}(\omega) - S_{V}(-\omega) =
2\r0\hbar\omega.\label{ResistorDensityAsymm} \ee This quantum
treatment can also be applied to any arbitrary dissipative
network\cite{DevoretLesHouches}. If we have a more complex circuit
containing capacitors and inductors, then in all of the above
expressions, $\r0$ should be replaced by ${\rm Re} Z(\omega)$
where $Z(\omega)$ is the complex impedance presented to the qubit.

\section{The Single Cooper-Pair Box: a Two-Level Quantum Circuit\label{SecBoxBasics}}

\begin{figure}
\includegraphics{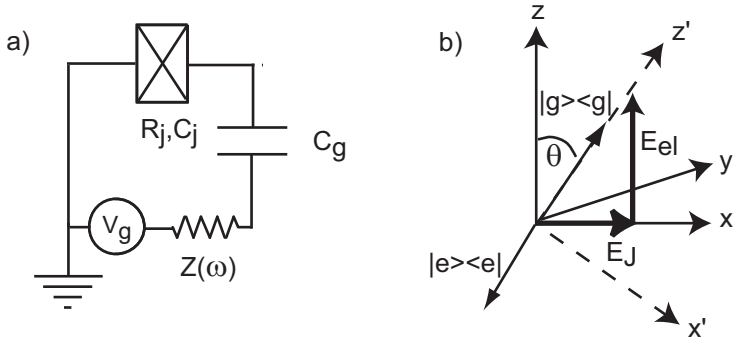}\caption{a) Circuit diagram of Cooper-pair box.
b) Pseudo-spin representation of the energies of Cooper-pair box.
The density matrix for the two pure eigenstates lie along the
total effective field, collinear with the z' axis.}
\label{CircuitAndBlochFig}\end{figure}

The Cooper-pair box (CPB) is a simple circuit \cite{Bouchiat},
consisting of a small superconducting ``island", connected to a
large reservoir via a single small-capacitance Josephson junction,
depicted as a box with a cross (Fig.~\ref{CircuitAndBlochFig}).
The island is charge biased by applying a voltage ($V_g$) to a
nearby lead, called the gate, which has a small capacitance to the
island, $C_g$. The junction is characterized by its capacitance,
$C_j$, and its tunnel resistance, $R_j$. At temperatures well
below the transition temperature of the superconductor ($T_C \sim
1.5$~K for the usual Al/AlOx/Al junctions), none of the many
($\sim 10^9$) quasiparticle states on the island should be
thermally occupied, and the number of Cooper-pairs on the island
is the only relevant degree of freedom.

We may then write the Hamiltonian for the box in terms of the
states of different numbers of pairs on the island, which are
eigenstates of the number operator, $\hat{n}\ketn=n\ketn$. The box
Hamiltonian consists of an electrostatic term, plus a Josephson
term describing the coupling of the island to the lead,
\begin{eqnarray}
H & = & H_{electrostatic} + H_{Josephson} \\
& = & 4E_C \sum_{n}(n-n_g)^2\ketn\bran - \frac{E_J}{2} \sum_{n}
(|n+1\rangle\langle n| + h.c.)
\end{eqnarray}
The energy scale for the electrostatic interaction is given by the
charging energy, $E_C = e^2/2C_\Sigma$, where $C_\Sigma= C_j+C_g$
is the total island capacitance, while the Josephson energy,
$E_J$, is set by the tunnel resistance and the gap of the
superconductor, \be E_J = \frac{h\Delta}{8e^2 R_j} =
\frac{\Delta}{8}\frac{R_K}{R_j}. \ee The electrostatic term is
easily modulated by changing the voltage on the gate; the quantity
$n_g = C_gV_g/2e$ that appears in the Hamiltonian corresponds to
the total polarization charge (in units of Cooper pairs) injected
into the island by the voltage source.

This Hamiltonian leads to particularly simple behavior in the
charge regime, when the electrostatic energy dominates over the
Josephson coupling, $4E_C \gg E_J$. In this case we can restrict
the discussion to only two charge states, $|n=0\rangle$ and
$|n=1\rangle$. For convenience we can reference the energies of
the two states to their midpoint, $E_{mid} = 4E_C(1-2n_g)^2$, so
that the Hamiltonian now becomes \be
    H  = \frac{1}{2}\left(
        \begin{array}{cc}
          -E_{el} & -E_J \\
          -E_J & E_{el} \\
        \end{array}
    \right)
\ee where $E_{el}$ is the electrostatic energy that is now {\em
linear} in the gate charge, $E_{el}= 4E_C(1-2n_g)$.  It is also
now apparent that the Hamiltonian is identical to that of a
fictitious spin-1/2 particle, \be
H=-\frac{E_{el}}{2}\sigma_z-\frac{E_J}{2}\sigma_x,\label{spinHamiltonian}\ee
under the influence of two psuedo-magnetic fields, $B_z = E_{el}$
and $B_x=E_J$, as depicted in Fig.~\ref{CircuitAndBlochFig}. In
other words, the box is a qubit or two-level system\footnote{Of
course, this is an approximation, as there are other charge states
($|n=2\rangle$, etc.) which are possible, but require much higher
energy. Even outside the charge regime (i.e. when $E_J \geq 4E_C$)
the two lowest levels of the box can be used to realize a qubit
\cite{Vion}. In this case, the two states do not exactly
correspond with eigenstates of charge, and matrix elements are
more complicated to calculate. Nonetheless, this regime can also
be used as an electrical quantum spectrum analyzer.}. The state of
the system is in general a linear combination of the states
$|n=0\rangle$ and $|n=1\rangle$. The state can be depicted using
the density matrix, which corresponds to a point on the Bloch
sphere, where the north pole (+z-direction) corresponds to
$|n=0\rangle$. The ground and excited states of the system will be
aligned and anti-aligned with the total fictitious field, i.e. in
the $\pm z'$ directions.

\begin{figure}
\includegraphics{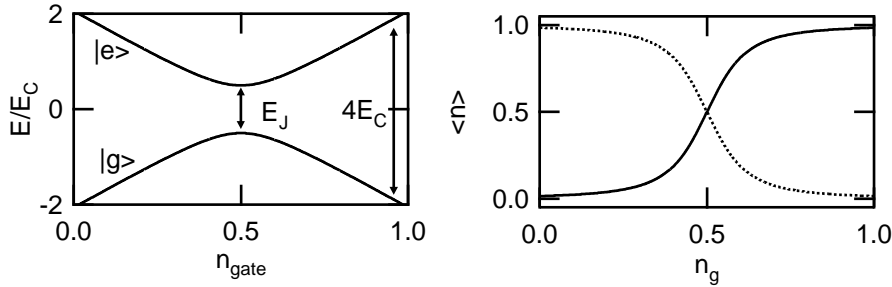}
\caption{Energies (left) of ground and excited states of a
Cooper-pair box with $E_C=E_J$ vs. dimensionless gate charge,
$n_g=C_gV_g/2e$.  The expectation value of a charge measurement,
$\langle n\rangle$, (right) for the ground (solid line) and
excited (dotted line) states vs. $n_g$.}
\label{EnergyAndChargeFig}\end{figure}

It is also apparent from this discussion that the states of the
box can be easily manipulated by changing the gate voltage. The
energies of the ground and excited states, as a function of $n_g$,
are displayed in Figure~\ref{EnergyAndChargeFig}. The energy
difference between the ground and excited bands varies from $4E_C$
at $n_g = 0,1$, to a minimum at the charge degeneracy point,
$n_g=1/2$. At this point, the Josephson coupling leads to an
avoided crossing, and the splitting is $E_J$.

Also plotted is the expectation value of the number operator,
$\langle|\hat{n}|\rangle$, which is proportional to the total
charge on the island. In the geometrical picture of
Fig.~\ref{CircuitAndBlochFig}b, a measurement of charge
($\hat{n}$) is equivalent to projecting the state on the z-axis,
$\hat{n}=\frac{1}{2}(1-\sigma_z)$. We see that as the gate charge
is changed from 0 to 1, the ground state is initially
$|n=0\rangle$, and the character of the ground and excited states
interchange on passing through the degeneracy point, leading to
the transition between $\langle \hat{n}\rangle$ = 0 and 1, which
is broadened by quantum fluctuations (the $\sigma_x$ coupling). At
the degeneracy point, the ground and excited states lie in the
$\pm$x-directions, i.e. they are symmetric and antisymmetric
combinations of the two charge states. In general, we will denote
the ground and excited state of the CPB at a particular gate
voltage as $\ketg$ and $\kete$, which are given in terms of the
charge states by
$\ketg=\cos(\theta/2)|0\rangle+\sin(\theta/2)|1\rangle$ and
$\kete=-\sin(\theta/2)|0\rangle+\cos(\theta/2)|1\rangle$
respectively, where $\theta= \arctan[E_J/E_{el}]$ is a function of
the gate voltage.

A nice property of the CPB in this regime is that the various
matrix elements can be calculated in a straightforward way. For
example, the expectation value of $\hat{n}$ in the ground state,
$\brag\hat{n}\ketg$, is therefore equal to
$1/2(1-\brag\sigma_z\ketg)=\sin^2(\theta/2)$, from which we can
find the ground state charge as shown in
Fig.~\ref{EnergyAndChargeFig}. A perturbation in the gate charge,
due for example to a fluctuation or change in the applied gate
voltage, will lead to a proportional change in the electrostatic
energy, or the z-component of the fictitious magnetic field. Such
a perturbation will cause both dephasing and transitions between
states.

\section{General Discussion of CPB Coupled to a Dissipative Environment\label{SecCPBPlusEnv}}

In the previous section we described the Hamiltonian and the
eigenstates for a Cooper-pair box which is ``charge-biased," i.e.
controlled with a voltage applied to a gate capacitor $C_g$, as
shown in Fig.~\ref{CircuitAndBlochFig}. In our earlier treatment
of the box, the voltage and the dimensionless gate charge, $n_g$
were treated as fixed parameters of the Hamiltonian (c-numbers).
In this case, the box's evolution is purely deterministic and
conservative. However, it is impossible, even in principle, to
control such a voltage with arbitrary precision at all
frequencies. In Fig.~\ref{CircuitAndBlochFig}, the idealized
source of the gate voltage is drawn in series with an impedance
$Z(\omega)$ of the gate lead. Generally, this gate lead will be
connected to external wiring (a transmission line), with a typical
real impedance comparable to the impedance of free space ($\sim
50~\Omega$) at the microwave transition frequencies of the box.
>From the fluctuation-dissipation theorem we know that this
impedance will introduce noise on the gate voltage, even at zero
temperature.

There are several effects of the voltage noise on the box, or the
coupling of our spin-1/2 circuit to the many external degrees of
freedom represented by the gate impedance. First, even at zero
temperature, we will find a finite excited state lifetime, $T1$,
for the box. Second, at finite temperature, we will find a finite
polarization of our psuedo-spin, i.e. some steady-state
probability to find the spin in its excited state. Finally, the
gate noise introduces a random effective field felt by the spin,
and a loss of phase coherence for a superpostition state. It is
this last effect which is most important in making high-fidelity
qubits and performing quantum computations, but it is the first
two which depend most explicitly on the quantum nature of the
noise. We deal in this manuscript with only these first two
features of the box's coupling to the electromagnetic environment,
and ignore the dephasing\footnote{For a nice recent treatment of
dephasing in Josephson junctions, see Ref. \cite{Martinis}.}. Of
course, the other parameter in the Hamiltonian, the Josephson
energy, can in principle fluctuate, especially as in many
experiments the box's junction is split into a small SQUID in
order to provide external tuning of $E_J$ with an applied flux. We
concentrate here only on the voltage noise (fluctuations in the
$\sigma_z$ part of the Hamiltonian) for simplicity.

We begin with a very simple treatment of the dynamics of the
two-level system under the influence of the gate voltage noise. We
are interested in the ensemble-averaged behavior of our
psuedo-spin, which is best described using the density matrix
approach, and is detailed in Section~\ref{SETsection} on the
coupling of the box to a measuring SET. The basic effects, apart
from dephasing, however, can be captured simply by examining the
probabilities $p_g$ and $p_e$ of finding the box in its ground
($\ketg$) or excited ($\kete$) states. The noise of the external
environment can drive transitions from ground to excited state and
vice-versa, at rates $\gammaup$ and $\gammadown$, respectively.
The coupled master equations for these probabilities are
\begin{eqnarray} \frac{{dp_e }} {{dt}} &=
p_g \Gamma _ \uparrow   - p_e \Gamma _ \downarrow \label{masterpground}\\
\frac{{dp_g }} {{dt}} &= p_e \Gamma _ \downarrow   - p_g \Gamma _
\uparrow \label{masterpexcited}\end{eqnarray} Of course
conservation of probability tells us that $p_e+p_g=1$, so we
introduce the polarization of the spin-1/2 system, $P=p_g-p_e$. In
steady-state, the detailed balance condition is  $p_e\gammadown =
p_g\gammaup$. The two rates $\gammaup$ and $\gammadown$ are
related by Equations \ref{gamup} and \ref{gamdown} to the spectral
densities of the noise at negative and positive frequencies. We
see immediately that if the spectral density is symmetric
(classical!), then the rates for transitions up and down are
equal, the occupancies of the two states are exactly equal, and
the polarization of the psuedo-spin is identically zero. It is the
quantum, or antisymmetric, part of the noise which gives the
finite polarization of the spin. Even in NMR, where the
temperature is large compared to the level splitting
$(\hbar\omega_{01}\leq\kbt)$, this effect is well-known and
crucial, as the small but non-zero polarization is the subject of
the field!

Solving for the steady-state polarization, we find \be
P_{ss}=\frac{\gammadown-\gammaup}{\gammadown+\gammaup}=\frac{S(+\w01)-S(-\w01)}{S(+\w01)+S(-\w01)}
\label{PolSteadyStateEq} \ee An measurement of the steady-state
polarization allows one to observe the amount of asymmetry in the
noise, so the two-level system is a {\em quantum} spectrum
analyzer.

If we can create a non-equilibrium polarization, $P=P_{ss}+\Delta
P$ (a pure state is not necessary) of our two-level system, we
expect it to return to the steady state value. Substituting the
modified probabilities $p_e(t)=p_{e_{ss}}-\Delta P(t)/2$ and
$p_g(t)=p_{g_{ss}}+\Delta P(t)/2$ into our master equations above,
we find an equation for the deviation of the polarization \be
\frac{d(\Delta P (t))}{dt}=-\Delta P(t)(\gammaup+\gammadown)
\label{masterpol}.\ee Thus the system decays to its steady-state
polarization with the relaxation rate
$\Gamma_1=\gammaup+\gammadown = (A/\hbar)^2[S(-\w01)+S(+\w01)]$
related to the {\it total} noise at both positive and negative
frequencies. In NMR, the time $1/\Gamma_1$ is referred to as $T1$.
In the zero-temperature limit, there is no possibility of the
qubit absorbing energy from the environment, so $\gammaup=0$, and
we find full polarization $P=1$, and a decay of any excited state
population at a rate $\gammadown=1/T1$ which is the spontaneous
emission rate.

It is worth emphasizing that a quantum noise source is always
characterized by two numbers (at any frequency), related to the
positive and negative frequency spectral densities, or to the
symmetric and antisymmetric parts of the noise. These two
quantities have different effects on a two-level system,
introducing a finite polarization and finite excited-state
lifetime. Consequently, a measurement of {\em both} the
polarization and $T1$ of a two-level system is needed to fully
characterize the quantum noise coupled to the qubit. Such
measurements in electrical systems are now possible, and some of
us \cite{Lehnert} have recently performed such a characterization
for the specific case of a CPB coupled to a superconducting
single-electron transistor.

\begin{table}
\caption{Different ways to characterize a quantum
reservoir.}\label{tab1}
\begin{tabular}{lcc}\hline\hline
Fermi Golden Rule &  $\Gamma _ \uparrow  (\omega) = \frac{{A^2 }}
{{\hbar^2 }}S_V ( + |\omega |)$ & $\Gamma _ \downarrow  (\omega)
= \frac{{A^2 }} {{\hbar ^2 }}S_V ( - |\omega |)$ \\
\hline Fluct.-Diss. Relation &  $n_\gamma (\omega ) = 2\Gamma _
\uparrow / (\Gamma _ \downarrow   - \Gamma _ \uparrow )$ &
${\textrm Re} [Z(\omega )] = \frac{\hbar }{A^2 \omega}(\Gamma _
\downarrow   - \Gamma _ \uparrow
)$ \\
\hline NMR & $ T1 = (\Gamma _ \downarrow   + \Gamma _
\uparrow)^{-1} $ & $P = (\Gamma _ \downarrow   - \Gamma _
\uparrow) /(\Gamma
_ \downarrow   + \Gamma _ \uparrow )$ \\
\hline Quantum Optics & $B_{Einstein}  = \Gamma _ \uparrow$ &
$A_{Einstein}
=\Gamma _ \downarrow   - \Gamma _ \uparrow$\\
\hline
\end{tabular}
\end{table}

Our discussion in this section uses the language of NMR to
describe the effects on the two-level system. There are, however,
several possible protocols\footnote{The idea of watching the decay
from the pure states $\kete$ and $\ketg$ to measure
$S_V(\pm|\w01|)$ separately was described in
Section~\ref{TLSanalyzer}.} for measuring the quantum noise, and
several different ``basis sets" or measured quantities which
describe the noise process or the quantum reservoir to which the
two-level system is coupled. Table~1 contains a ``translation"
between the specific pairs of quantities that are commonly used in
different disciplines, and their relation to the positive and
negative noise spectral densities. In all cases, though, {\em two
separate numbers} are required to specify the properties of a
quantum reservoir.

\section{The Box Coupled to an Ohmic Environment \label{BoxOhmicSection}}

We can now proceed to the effects of a specific dissipative
coupling to the Cooper-pair box. The noise on the gate voltage
will lead to a fluctuation of the gate charge parameter, $n_g$,
and thus to a fluctuation of the electrostatic energy, i.e. the
$\sigma_z$ term in the Hamiltonian (Eq.~\ref{spinHamiltonian}).
Depending on the average value of $n_g$, this fluctuation will
consist of fluctuations which are both longitudinal ($\parallel$
to $\sigma'_z$) and transverse ($\perp$ to $\sigma'_z$). To
calculate the rates of transitions between the states $\kete$ and
$\ketg$, we need to find the coupling strength $A$ of this
perturbation in the $\sigma'_x$ direction. Referring to
Fig.~\ref{CircuitAndBlochFig}, we see that $\sigma_z=\cos
(\theta)\sigma'_z-\sin(\theta)\sigma'_x$. If we let the gate
charge now be $n_g(t) = \bar{n_g} + \delta n_g(t)$, we may rewrite
the Hamiltonian Eq.~\ref{spinHamiltonian} in the new eigenbasis as
\be H = -\frac{E_{01}}{2\hbar}\sigma'_z + 4E_C\cos(\theta)\delta
n_g(t)\sigma'_z - 4E_C\sin(\theta)\delta
n_g(t)\sigma'_x~.\label{spinHrotated}\ee The time-varying term in
the $\sigma'_z$ direction will effectively modulate the transition
frequency, $\w01=E_{01}/\hbar$, and cause dephasing. In terms of
the gate voltage noise, $V(t)$, the $\sigma'_x$ perturbation term
has the form $AV(t)\sigma'_x=e\kappa\sin(\theta)V(t)\sigma'_x$,
where $e$ is the electron's charge and $\kappa=C_g/C_\Sigma$ is
the capacitive coupling. Using Eq.~\ref{gamdown}, we find \be
\gammadown=\left(\frac{e}{\hbar}\right)^2\kappa^2\sin^2(\theta)S_V(+\w01)\label{CPBrelaxrate}.\ee
If the environment is effectively at zero temperature
($\hbar\w01\gg\kbt$), then $S_V(+\w01)=2\hbar\w01 R$, and the
quality factor of the transition is \be
Q=\w01/\gammadown=\frac{1}{\kappa^2\sin^2\theta}\frac{R_K}{4\pi
R}~,\label{qfactor}\ee where $R_K=h/e^2$ is the resistance
quantum.

\begin{figure}
\includegraphics{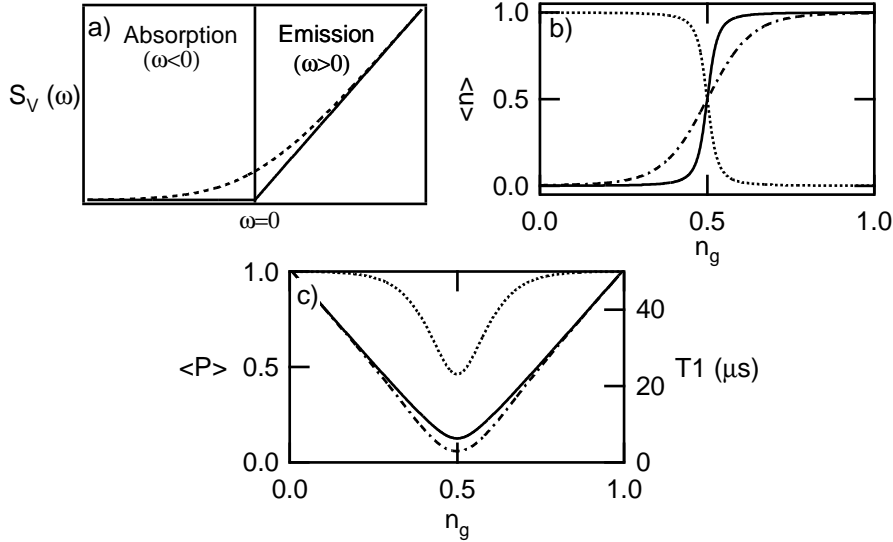}
\caption{The box coupled to an equilibrium, Ohmic environment,
i.e. a resistor. a) Two-sided noise spectral density, of the
voltage, $S_V(\omega)$, for a resistor at T=0 (solid line) and
finite temperature (dashed line) b) Average charge of box with
$E_C = 1 K$, $E_J=0.5 K$ when coupled with strength $\kappa=0.01$
to a resistor with resistance $R=50~\Omega$ and temperature $
T=0.5K$. c) Polarization (dotted line) and relaxation time T1 for
the same parameters. Full line is the rate of spontaneous
emission, i.e. T1 at zero temperature. } \label{OhmicFig}
\end{figure}

For a finite temperature, we have rates in both directions, and
the polarization of the psuedo-spin is given by the ratio of the
antisymmetric (Eq.~\ref{ResistorDensityAsymm}) to symmetric
(Eq.~\ref{ResistorDensitySymm}) spectral densities,
$P=\tanh(\hbar\w01/2\kbt)$, as one expects for any two-level
system at temperature $T$. An example of the average charge state
of a Cooper-pair box at finite temperature, and of the
polarization and equilibration time T1, are shown in
Figure~\ref{OhmicFig}. As the gate voltage is varied, the
transition frequency of the box changes from a maximum near
$n_g=0$, to a minimum $\w01=E_J/\hbar$ at the degeneracy point
$n_g=0.5$ and then back again. We see that the states of the box
are generally most ``fragile" near the avoided crossing. First,
the energy splitting is a minimum here, leading to the lowest
polarization of the psuedo-spin. Second, because the eigenstates
$\ketg$ and $\kete$ point in the $\sigma_x$ direction, the matrix
elements for the voltage fluctuations of the environment are
maximal, i.e. the noise is orthogonal to the spin. This also
implies that the dephasing effects are minimal at this degeneracy
point, which offers great advantages for improving the decoherence
times \cite{Vion}, but the excited state lifetimes are smallest at
this point. One also sees that the lifetimes become large away
from the degeneracy, where the electrostatic energy dominates over
the Josephson energy, which offers advantages when measuring the
charge state. In the limit that $E_J$ could be suppressed to zero
during a measurement, the matrix elements (for voltage noise)
vanish, and a quantum non-demolition (QND) measurement
\cite{Braginsky} could be performed. The idea of using the qubit
as a quantum spectrum analyzer is precisely the reverse, where we
measure the ``destruction" in the two-level system (i.e. inelastic
transitions caused by the coupling of the states to the
environment), in order to learn about the quantum noise spectrum
of the fluctuations.

\begin{figure}
\includegraphics{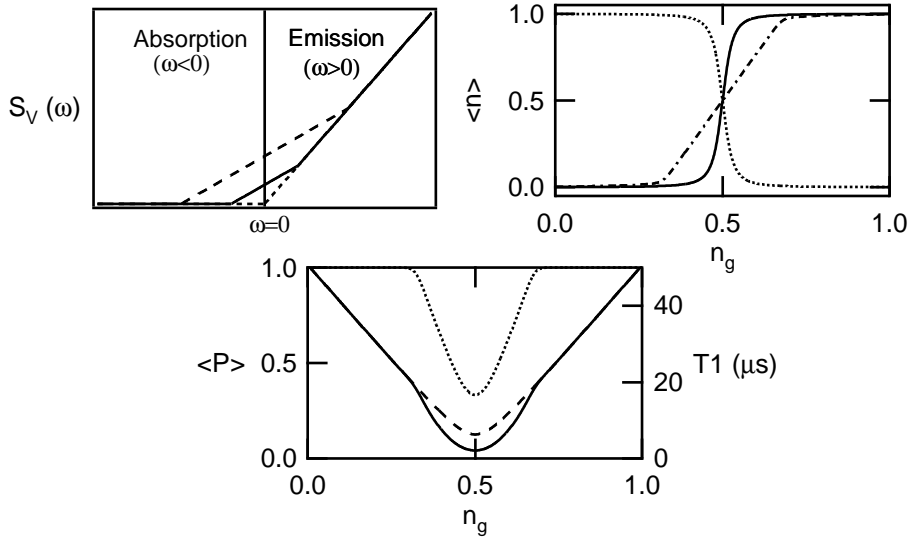}
\caption{The box coupled to an nonequilibrium, Ohmic environment,
i.e. a 50~$\Omega$ tunnel junction. a) Two-sided noise spectral
density, of the voltage, $S_V(\omega)$, for a junction at T=0.02
K, with zero voltage (dotted line), and increasing bias voltages
(solid and dashed lines). b) Average charge of box with $E_C = 1
K$, $E_J=0.5 K$ when coupled with strength $\kappa=0.01$ to a
junction biased at $eV=1.5$~K. c) Polarization (dotted line) and
relaxation time T1 for the same parameters (solid) and for T=0,
V=0 (dashed line).} \label{ShotFig}\end{figure}

The Cooper-pair box can of course also be used to measure the more
interesting spectral densities of {\em nonequilibrium} devices.
The simplest example is to replace the gate resistance by a tunnel
junction. If we arrange to bias the junction using, e.g. an
inductor, a dc current $I$ and an average dc voltage $V$ can be
maintained across the junction. Classically, the current noise of
such a tunneling process is frequency independent, $S_I = 2eI$.
The voltage noise density presented to the CPB's gate would then
be $S_V=2eIR^2_T$, where $R_T$ is the junction tunnel resistance.
In fact, this ``white" spectral density can only extend up to
frequencies of order $\omega=eV/\hbar$, the maximum energy of
electrons tunneling through the junction. The correct
high-frequency form of the {\em symmetrized} noise density was
calculated by Rogovin and Scalapino \cite{RogScal}, \be S_V \left(
\omega \right)   = R\left( {\hbar \omega + eV} \right)\coth \left[
{\frac{{\hbar \omega  + eV}} {{2k_B T}}} \right] + R\left( {\hbar
\omega  - eV} \right)\coth \left[ {\frac{{\hbar \omega  - eV}}
{{2k_B T}}} \right] \label{ShotSymm},\ee and was {\em indirectly}
measured in a mesoscopic conductor using a conventional spectrum
analyzer \cite{ShotNoisePRL}. This noise can also be expressed
\cite{Aguado} in its two-sided form \be S_V \left( \omega \right)
= \frac{{\left( {\hbar \omega + eV} \right)R_T }} {{1 - e^{ -
\frac{{\hbar \omega + eV}} {{k_B T}}} }} + \frac{{\left( {\hbar
\omega  - eV} \right)R_T }} {{1 - e^{ - \frac{{\hbar \omega - eV}}
{{k_B T}}} }} \label{ShotTwoSided},\ee and is displayed in
Fig.~\ref{ShotFig}. Notice that the antisymmetric part of this
noise is the same as that of the ordinary resistor,
$S_V(+\omega)-S_V(-\omega)=2\hbar\omega R_T$, and is independent
of the voltage. Also shown in Fig.~5 is the polarization and
relaxation time, T1, of a CPB coupled to a shot noise environment.
We see that full polarization is achieved only when $\hbar\w01\gg
eV$. For low transition frequencies, the polarization is inversely
proportional to the current through the junction. Aguado and
Kouwenhoven \cite{Aguado} have described the use of a double
quantum dot as a two-level system to probe this behavior of the
shot noise.

Given our discussion so far, it is now interesting to ask what the
effects of a real quantum measurement on a quantum circuit will
be. A quantum measurement device will in general be neither
linear, Ohmic, nor equilibrium. Obviously, if we hope to
characterize this measurement process, and to understand what one
will observed when the qubit is coupled to the noise processes of
the measuring device, we will need to calculate the full quantum
(two-sided!) noise spectrum of the amplifier or detector.

\section{Single-Electron Transistor Coupled to a Two-Level
System \label{SETsection}}

We have seen in previous sections that a two-level system (TLS)
may be used as a ``spectrum analyzer" to measure quantum noise.
Here, we use this technique to {\it theoretically} calculate
quantum noise.  Instead of simply studying the ``noisy" system of
interest in isolation, one can study a composite system consisting
of the ``noisy" system coupled to a TLS; by calculating the
relaxation and excitation rates of the TLS, one can efficiently
calculate the quantum noise of interest\footnote{Note that the
spirit of our approach is similar to that employed in the theory
of full counting statistics \cite{Levitov}.  There too one
attaches an auxiliary spin 1/2 to the scattering system of
interest, and studies the dynamics of the fully coupled system to
obtain the statistics of charge transfer in the scatterer.}. We
demonstrate the usefulness of this technique by outlining a
calculation of the quantum charge noise of a single electron
transistor (SET).  This is an important example, as when an SET is
used as an electrometer, it is this noise which determines the
measurement backaction.

The SET consists of a metallic island attached via tunnel
junctions to source and drain reservoirs.  It is described by the
Hamiltonian:
\begin{eqnarray}
    H_{SET} & = &
        \sum_{k,\alpha=L,R,I} \left(\ve_{k} - \mu_{\alpha} \right)
            c^{\dagger}_{k \alpha}
            c^{\phantom{\dagger}}_{k \alpha}    +
        E_C ( n - \NN)^2 + H_T \\
    H_T & = &
        t \sum_{k,q, \alpha=L,R} \left(
            F^{\dagger}
            c^{\dagger}_{k I}
            c^{\phantom{\dagger}}_{q \alpha} +  h.c. \right)
\end{eqnarray}
The first term in $H_{SET}$ describes the kinetic energy of
electrons in the leads ($\alpha = L,R$) and on the island ($\alpha
= I$).  The second term is the Coulomb charging energy which
depends on $n$, the number of excess electrons on the island. This
interaction term can be tuned by changing the voltage on a nearby
gate electrode which is capacitively coupled to the island; $\NN$
represents the dimensionless value of this voltage. Finally, $H_T$
describes the tunneling of electrons through the two SET tunnel
junctions; the conductance of each junction (in units of $e^2/h$)
is given by $g = 4 \pi^2 t^2 \nu_0^2$, with $\nu_0$ being the
density of states.  $F^{\dagger}$ is an auxiliary operator which
increases $n$ by one: $ \left[n, F^{\dagger} \right] =
F^{\dagger}$.  For simplicity, we assume that the two junctions of
the SET are completely symmetric (i.e. equal junction capacitances
and dimensionless conductances).

Throughout this section, we will be interested in the regime of
sequential tunneling in the SET, where transport involves
energy-conserving transitions between two charge states of the SET
island, say $n=0$ and $n=1$.  These transitions are described by
simple rates, which can be derived via Fermi's Golden rule:
\begin{eqnarray}
    \Gamma_{n\pm1,n}^{\alpha}  & = &
        \gamma( \left[\Delta E\right]^{\alpha}_{n\pm1,n} )
        \label{QPRates}\\
    \gamma(\Delta E) & = &
    \frac{g \Delta E / h}{1 - e^{- \Delta E / (k_B T) } }
     \label{gammadefn}\\
    \Delta E^{\alpha}_{n\pm1,n} & = & \mp
        2 E_C \left(n \pm \frac{1}{2} - \NN\right)
         \pm \left(\frac{1}{2} - \delta_{\alpha,R} \right)e V_{DS}
         \label{DeltaE}
\end{eqnarray}
$\Gamma_{n \pm 1,n}^{\alpha}$ is the tunneling rate from the
charge state $n$ to $n \pm 1$ through junction $\alpha$; $\Delta
E$ is the energy gained in making the tunneling transition, and
includes contributions both from the drain-source voltage $V_{DS}$
and from the charging energy.  Sequential tunneling is the
dominant transport mechanism when the junction conductances are
small (i.e. $g/(2 \pi) \ll 1$), and the dimensionless gate voltage
$\NN$ is not too far away from a charge degeneracy point.
Sequential tunneling is the most important regime for measurement
applications, as it yields the largest SET currents.

At low temperatures, only tunnel events which follow the voltage
are possible.  There are thus there only two relevant rates: $n=0
\ra 1$ transitions occur through the left junction at a rate
$\Gamma_{10}^{L}$, while $n=1 \ra 0$ occur through the right
junction at a rate $\Gamma_{01}^{R}$. The average current will be
given by:
\begin{equation}
    \langle I \rangle = e \bar{\Gamma} \equiv e \frac{ \Gamma^L_{10} \Gamma^R_{01} }
        {  \Gamma^L_{10} + \Gamma^R_{01} }
\end{equation}

We are interested in calculating $S_Q(\omega)$, the quantum noise
associated with fluctuations of the charge on the central island
of the SET.  It is defined as:
\begin{equation}
    S_Q(\omega) = \int_{-\infty}^{\infty} dt \langle
        n(t) n(0) \rangle e^{-i \omega t}
\end{equation}
Note that we can equivalently think of $S_Q$ as describing the
voltage noise of the island, as $V_{island} = e n / C_{\Sigma}$,
where $C_{\Sigma}$ is the total capacitance of the island.  In two
limits, the form of the island charge noise can be anticipated.
For $\omega \ra 0$, the noise will correspond to classical
telegraph noise-- the island charge $n$ fluctuates between the
values $0$ and $1$, with Poisson-distributed waiting times
determined by the rates $\Gamma^L_{10}$ and $\Gamma^R_{01}$. We
thus expect a symmetric, Lorentzian form \cite{NatureBackaction}
for the noise at low frequencies:
\begin{equation}
    S_Q(\omega) \ra \frac{2 \bar{\Gamma} }{\omega^2 +
        (\Gamma^L_{10} + \Gamma^R_{01})^2 }
        \hspace{1 cm} (\omega \ll E_C)
\label{ClassicalSQ}
\end{equation}
For large frequencies $|\omega| \gg E_C$, we expect that
correlations due to the charging energy will have no influence on
the noise.  The system will effectively look like two tunnel
junctions in parallel, and we can use the results of
Sec.~\ref{BoxOhmicSection} for the corresponding voltage noise.
Noting that each junction effectively consists of a resistor and
capacitor in parallel, we have at zero temperature:
\begin{eqnarray}
    S_Q(\omega) & = & \frac{C_{\Sigma}^2}{e^2} \times S_V(\omega)
         \ra
         \frac{C_{\Sigma}^2}{e^2}  \left[
            2 \hbar \omega
            \textrm{Re } Z_{tot} \Theta(\omega)
         \right]
        \hspace{0.5 cm} (|\omega| \gg E_C) \nonumber \\
        & = & 4 \left(\frac{g}{2 \pi}\right)  \frac{  \omega }
            { \left( \frac{g}{2 \pi} \frac{4 E_C}{\hbar} \right)^2
            + \omega^2 } \Theta(\omega)
            \label{Nyquist}
\end{eqnarray}
Note that $S_Q(\omega)$ decays as $1/\omega$ at large frequencies,
whereas Eq. (\ref{ClassicalSQ}) for classical telegraph noise
decays as $1/\omega^2$.

Given these two limiting forms, the question now becomes one of
how the SET interpolates between them.  One might expect that the
two results should simply be added in quadrature, but as with
combining thermal and quantum noise (see
Sec.~\ref{BoxOhmicSection}), this is approach is too simple.  A
completely quantum mechanical way of calculating the noise for any
frequency is needed. This was recently accomplished by Johansson
{\it et.~al} \cite{Wendin} using an extension of a technique
developed by Sch\"{o}ller and Sch\"{o}n \cite{Schoeller}. Here, we
re-derive their results using the coupled system approach outlined
above.  This method is physically motivated and allows for a
heuristic interpretation of the final result.

\section{SET Coupled to a Qubit}

We now consider a system where the SET is coupled to a two-level
system (i.e. a qubit), with a coupling Hamiltonian which can
induce transitions in the TLS.  Using spin operators to describe
the qubit, and assuming operation at the degeneracy point for
simplicity, where the transitions are fastest\footnote{This
amounts to maximizing the ``destruction" due to the SET's noise,
and the case where $\theta=\pi/2$, the qubit eigenstates are in
the $\sigma'_z=\sigma_x$ direction, and the SET's perturbation is
in the $-\sigma'_x=\sigma_z$ direction (c.f.
Eq.~\ref{spinHrotated}). After the noise of the SET is found, we
can then recalculate the effects on the qubit at various $n_g$ or
values of $\theta$ by including the modified matrix elements in
the coupling coefficient, $A$.}, we have:
\begin{equation}
    H = H_{SET} -\frac{1}{2} \Omega \sigma_x + A \sigma_z
    n, \label{}
\end{equation}
where $\Omega$ is the qubit splitting
frequency\footnote{Henceforth we use $\Omega$ for the transition
frequency, instead of the previous notation $\w01=E_{01}/\hbar$,
for compactness.}, and $A$ is the coupling strength. We can define
the rates $\gammaup$ and $\gammadown$ which are, respectively, the
rate at which the qubit is excited by the SET, and the rate at
which the qubit is relaxed by the SET. For a weak coupling ($A \ra
0$), one has (c.f. Eq.~\ref{gamdown},\ref{gamup} in
Sec.~\ref{TLSanalyzer}):
\begin{equation}
    \Gamma_{\downarrow / \uparrow} = \frac{A^2}{\hbar} S_Q(\pm \Omega)
\label{FGR}
\end{equation}
Eq. (\ref{FGR}) tells us that if we know the rates $\gammaup$ and
$\gammadown$ for a weakly coupled system at an arbitrary splitting
frequency $\Omega$, we know the quantum noise $S_Q(\Omega)$ at all
frequencies. This is the essence of the technique previously
described, in which a qubit acts as a quantum spectrum analyzer of
noise. Here, we mimic this approach theoretically by obtaining
$\gammaup$ and $\gammadown$ from a {\it direct} analysis of the
coupled system in the limit of weak coupling ($A \ra 0$).  The
object of interest is the reduced density matrix $\rho$ which
describes both the charge $n$ of the transistor island {\it and}
the state of the qubit. We are interested in two quantities.
First, what is the stationary state of the qubit? The stationary
populations of the two qubit states (which are determined from the
time-independent solution for $\rho$) will tell us the
polarization of the qubit, and the amount of asymmetry in the
noise (c.f. Eq.\ref{PolSteadyStateEq}). Second, how quickly do the
qubit populations relax to their stationary value?  This
relaxation will be described by a time-dependent solution of
$\rho$ characterized by a mixing rate $\Gamma_{mix}$ which is the
{\it sum} of $\gammaup$ and $\gammadown$ (c.f.
Eq.~\ref{masterpol}). From these two results we can solve for the
individual values of $\Gamma_{\uparrow/\downarrow}$.

To deal with the dynamics of $\rho$, we make use of the fact that
sequential tunneling processes are completely described by
lowest-order perturbation theory in the tunneling Hamiltonian
$H_T$. Keeping only second order terms (there are no non-vanishing
first order terms), one obtains the following standard evolution
equation in the interaction picture:
\begin{equation}
    \label{RhoEvol}
    \frac{d}{dt} \rho(t) =
        -\frac{1}{\hbar} \int_{-\infty}^{t} dt' \langle
            \left[
                H_T(t), \left[
                    H_T(t'),  \rho(t') \otimes \rho_{F}
                    \right]
                    \right]
                        \rangle
\end{equation}
The angular brackets denote the trace over the single-particle
degrees of freedom in the SET leads and island; $\rho_{F}$ is the
equilibrium density matrix corresponding to the state of these
degrees of freedom in the absence of tunneling.\footnote{ In the
diagrammatic language of Ref. \cite{Schoeller},
Eq.~(\ref{RhoEvol}) is equivalent to keeping all $(H_T)^2$ terms
in the self-energy of the Keldysh propagator governing the
evolution of $\rho$.}  Note that a similar density matrix analysis
of a qubit coupled to a SET was recently discussed by Makhlin {\it
et.~al} \cite{Makhlin}; unlike the present case, these authors
restricted attention to a vanishingly small splitting frequency
$\Omega$.

To make progress with Eq. (\ref{RhoEvol}), we make a Markov
approximation, which involves replacing $\rho(t')$ on the
right-hand side with $\rho(t)$.  This is permissible as we are
interested in the {\it slow} dynamics of $\rho$.  We want to find
both the stationary solution of $\rho$, for which the Markov
approximation is exact, and the mixing mode, a mode whose time
dependence is $\propto e^{- (\Gamma_{\ua} + \Gamma_{\da}) t}$.
This mode is also arbitrarily slow in the weak coupling limit $A
\ra 0$ of interest. Finding the stationary mode and the mixing
mode correspond to evaluating the polarization and T1 of the qubit
(c.f. Eq.~\ref{PolSteadyStateEq} and Eq.~\ref{masterpol}), as was
shown earlier for the master equation of the {\em probabilities}
in Section~\ref{SecCPBPlusEnv}. Note that the Markov approximation
should be made in the {\it Schr\"{o}dinger} picture, as it is in
the Schr\"{o}dinger picture that $\rho$ will be nearly stationary
(i.e. all oscillations associated with the qubit splitting
frequency $\Omega$ will be damped out in the long-time limit).

Evaluation of Eq. (\ref{RhoEvol}) in the Markov approximation
results in the appearance of rates which are generalizations of
those given in Eq. (\ref{QPRates}). Now, however, these rates
depend on the initial and final state of the qubit-- tunneling
transitions can simultaneously change both the charge state of the
SET island {\it and} the state of the qubit. The resulting
equation is most easily presented if we write the reduced density
matrix $\rho$ in the basis of eigenstates at zero tunneling. For
each value of island charge $n$, there is a different qubit
Hamiltonian, and correspondingly a different a qubit ground state
$|g_n \rangle$ and excited state $|e_n \rangle$.  When a tunneling
event occurs in the SET, there is a sudden change in the qubit
Hamiltonian.  As the qubit ground and excited states at different
values of $n$ are {\it not} orthogonal, tunneling transitions in
the SET are able to cause ``shake-up" transitions in the qubit. In
the limit $A \ra 0$, the relevant matrix overlaps are given by:
\begin{eqnarray}
    \langle g_m | g_n \rangle & = & 1 - \frac{1}{2}
            \left( \frac{ A (m-n) }{\Omega} \right)^2
            = \langle e_m | e_n \rangle \\
    \langle e_m | g_n \rangle & = &
             \frac{ A (m-n) }{\Omega} \label{EGOverlap}
\end{eqnarray}
Defining the frequency dependent rate $\Gamma_{n \pm 1,n}(\omega)$
as:
\begin{equation}
     \Gamma_{n \pm 1,n}(\omega)  \equiv
        \sum_{\alpha=L,R}
        \gamma( \left[\Delta E \right]^{\alpha}_{n \pm 1,n}
        + \hbar \omega   ),
        \label{OmegaRates}
\end{equation}
where $\Delta E$ and $\gamma(\Delta E)$ are defined in Eqs.
(\ref{DeltaE}) and (\ref{gammadefn}), the required tunnel rates
take the form:
\begin{eqnarray}
    \Gamma_{m,n}  \equiv  \Gamma_{m,n}(0)
    \hspace{0.8 cm}
    \Gamma_{m,n}^\pm  \equiv  \Gamma_{m,n}(\pm \Omega)
    \label{QPRates2}
\end{eqnarray}
The $\Gamma^{+}$ rates correspond to tunneling events where the
qubit is simultaneously relaxed, and thus there is an additional
energy $\Omega$ available for tunneling.  For large $\Omega$,
tunneling processes which are normally energetically forbidden can
occur if they are accompanied by qubit relaxation. Similarly, the
$\Gamma^{-}$ rates describe tunnelling events where the qubit is
simultaneously excited, with the consequence that there is less
energy available for tunneling.

Returning to the evolution equation Eq. (\ref{RhoEvol}), note that
we do not need to track elements of $\rho$ which are off-diagonal
in the island charge index $n$-- there is no coherence between
different charge states, as tunneling events necessarily create an
electron-hole excitation.  Further, if we focus on small qubit
frequencies, we may continue to restrict attention to only $n=0$
and $n=1$ (i.e. $\Omega$ is not large enough to ``turn on"
tunneling processes which are normally energetically forbidden).
Thus, there are 8 relevant matrix elements of $\rho$-- for each of
the four qubit density matrix elements (i.e. $g g$, $e e$, $g e$,
$e g$), there are two possible island charge states.  We combine
these elements into a vector $\vec{\rho} = \left(\rho_{g
g},\rho_{e e},\rho_{g e}, \rho_{e g} \right)$, where $\rho_{g g} =
\Big( \langle 0,g_0 | \rho | 0,g_0 \rangle , \langle 1,g_1 | \rho
| 1,g_1 \rangle \Big)$, etc. Organizing the resulting evolution
equation in powers of the coupling $A$, we obtain in the
Schr\"{o}dinger picture:
\begin{equation}
    \frac{d}{dt} \vec{\rho} = ( {\bf \Lambda_0} + \frac{A}{\Omega}
        {\bf \Lambda_1} +
        \frac{A^2}{\Omega^2} {\bf \Lambda_2} + ...) \vec{\rho}
    \label{SimpRhoEvol}
\end{equation}
We discuss the significance of the matrices ${\bf \Lambda_j}$ in
what follows.

The $8 \times 8$ matrix ${\bf \Lambda_0}$ describes the evolution
of the system at zero coupling:
\begin{eqnarray}
    \label{Lambda0}
    {\bf \Lambda_0} & = & \left(
    \begin{array}{cccc}
      M & 0 & 0 & 0 \\
      0 & M & 0 & 0 \\
      0 & 0 & +i \Omega  + M' & 0 \\
      0 & 0 & 0 & - i \Omega + M' \
    \end{array}
    \right),
\end{eqnarray}
with the $2 \times 2$ matrices $M$ and $M'$ being defined by:
\begin{eqnarray}
    M  =  \left(
        \begin{array}{cc}
          -\Gamma_{10} & \Gamma_{01} \\
          \Gamma_{10} & -\Gamma_{01} \\
        \end{array}
    \right)
    \hspace{0.8 cm}
    M'  =   \frac{1}{2} \left(
        \begin{array}{cc}
          -(\Gamma^+_{10} + \Gamma^-_{10}) & \Gamma^+_{01} + \Gamma^-_{01} \\
          \Gamma^+_{10} + \Gamma^-_{10}  & -(\Gamma^+_{01} + \Gamma^-_{01}) \\
        \end{array}
    \right)
\end{eqnarray}
At zero coupling there are no transitions between different qubit
states, and hence ${\bf \Lambda_0}$ has a block-diagonal form.
There are two independent stationary solutions of Eq.
(\ref{SimpRhoEvol}) at $A=0$ (i.e. two zero eigenvectors of ${\bf
\Lambda_0}$), which correspond to being either in the qubit ground
or qubit excited state:
\begin{eqnarray}
    \vec{z}_{g}  =
    \left(
        p_0,
        p_1,
        0,0,0,0,0,0 \right),
    \hspace{0.8 cm}
    \vec{z}_{e}  =  \left( 0,0,
        p_0,p_1,
        0,0,0,0 \right).
\end{eqnarray}
$(p_0,p_1)$ are the stationary probabilities of being in the $n=0$
or $n=1$ charge states:
\begin{equation}
    (p_0,p_1) = \left(
    \frac{ \Gamma_{01} }{ \Gamma_{01} + \Gamma_{10} },
        \frac{ \Gamma_{10} }{ \Gamma_{01} + \Gamma_{10} } \right)
\end{equation}
The existence of two zero-modes is directly related to the fact
that at zero coupling $(A=0)$, the probabilities to be in the
qubit ground and excited state are individually conserved.

At non-zero coupling, the matrices ${\bf \Lambda_1}$ and ${\bf
\Lambda_2}$ appearing in Eq. (\ref{SimpRhoEvol}) generate
transitions between different qubit states.  The matrix ${\bf
\Lambda_2}$ directly couples $\rho_{g g}$ and $\rho_{e e}$, while
${\bf \Lambda_1}$ couples $\rho_{g g}$ and $\rho_{e e}$ to the
off-diagonal blocks $\rho_{g e}$ and $\rho_{e g}$. The effect of
these matrices will be to break the degeneracy of the two zero
modes of Eq. (\ref{SimpRhoEvol}) existing at $A=0$. After this
degeneracy is broken, there will still be one zero mode $\rho_0$,
describing the stationary state of the {\it coupled} system (the
existence of a stationary solution is guaranteed by the
conservation of probability).  For weak coupling, the qubit
density matrix obtained from $\rho_0$ will be diagonal in the
basis $\{| g_{\langle n \rangle} \rangle, | e_{\langle n \rangle}
\rangle \} $, which corresponds to the average SET charge $\langle
n \rangle = p_1$.  The ratio of the occupancies of these two qubit
states will yield the ratio between the relaxation rate
$\gammadown$ and the excitation rate $\gammaup$.  In addition,
there will also be a slow, time-dependent mode of Eq.
(\ref{SimpRhoEvol}) arising from breaking the degeneracy of the
two $A=0$ zero modes. This time-dependent mode will describe how a
linear combination of $z_{g}$ and $z_{e}$ relaxes to the true
stationary state, and will have an eigenvalue $\lambda = -
\gammaup - \gammadown$, i.e. the mixing rate.

Thus, we need to do degenerate second order perturbation theory in
the coupling $A$ to obtain the relaxation and excitation rates
$\Gamma_{\da}$ and $\Gamma_{\ua}$. The only subtlety here is that
the matrix $M$ is not Hermitian, implying that it has distinct
right and left eigenvectors. Letting $\vec{\tilde{z}}$ represent
the left eigenvector of $\bf{ \Lambda_0 }$ corresponding to the
right eigenvector $\vec{z}$, we define the projector matrix ${\bf
P}$ as:
\begin{equation}
    {\bf P} = | z_{g} \rangle \langle \tilde{z}_{g} | +
            | z_{e} \rangle \langle \tilde{z}_{e} |,
\end{equation}
and let ${\bf P_{\bot}}$ denote $1 - {\bf P}$.  As usual,
degenerate second order perturbation theory requires diagonalizing
the perturbation in the space of the degenerate eigenvectors.  We
are thus led to look at the matrix $Q$, defined as:
\begin{equation}
    Q = \frac{A^2}{\Omega^2} \left(
        {\bf P \Lambda_2 P}  +
        \bf{ P \Lambda_1 P_{\bot} \left[ - \Lambda_0 \right]^{-1}
            P_{\bot} \Lambda_1 P } \right)
        \label{QDefn}
\end{equation}
>From the definition of Q, we may immediately identify the rates
$\gammaup$ and $\gammadown$:
\begin{eqnarray}
    \gammaup  =  \langle \tilde{z}_{e} | Q | z_{g} \rangle
    \label{UpDefn}
    \hspace{0.8 cm}
    \gammadown  =  \langle \tilde{z}_{g} | Q | z_{e} \rangle
    \label{DownDefn}
\end{eqnarray}

We thus see how the rates $\Gamma_{\ua,\da}$ arise in the present
approach-- they are related to breaking the degeneracy between two
zero-modes (stationary solutions) which exist at zero coupling.
Note that there are two distinct contributions to
$\Gamma_{\ua,\da}$, coming from the two terms in the matrix $Q$: a
``direct" contribution involving $\bf{ \Lambda_2}$ and an
``interference" contribution involving $\bf{ \Lambda_1}$ acting
twice. These two terms have a different physical interpretation,
as will become clear.

Let us first consider the rate $\gammaup$, which describes how
noise in the SET causes ground to excited state transitions in the
qubit.  For this rate, our approximation of only keeping two
charge states will be valid for all splitting frequencies
$\Omega$. To evaluate the ``direct" contribution to this rate,
which involves the first term in the matrix $Q$, note that the
relevant part of ${\bf \Lambda_2}$ has the expected form:
\begin{equation}
    {\bf \Lambda_2} | _{e e, g g}  = \left( \begin{array}{cc}
      0 & \Gamma^-_{01} \\
      \Gamma^-_{10} & 0 \
    \end{array}
    \right)
\end{equation}
i.e. it consists of tunnel rates which correspond to having given
up an energy $\Omega$ to the qubit.  Using Eqs. (\ref{QDefn}) and
(\ref{UpDefn}), we find:
\begin{equation}
    \gammaup |_{direct} =
        \left( \frac{A}{\Omega} \right)^2
        \left(
        p_0 \Gamma^-_{10} + p_1 \Gamma^-_{01} \right)
        = \left( \frac{A}{\Omega} \right)^2
        \left(
        \frac{\Gamma_{01} \Gamma^-_{10} + \Gamma_{10} \Gamma^-_{01}}
        { \Gamma_{10} + \Gamma_{01} } \right)  \label{Direct}
\end{equation}
The direct contribution to $\gammaup$ has a very simple form: for
each charge state $n=0,1$, add the rate to tunnel out of $n$ while
exciting the qubit, weighted by both the probability to be in
state $n$, and the overlap between ground and excited states (i.e.
$(A / \Omega)^2$). This is very similar to how one typically
calculates the current for a SET: one adds up the current
associated with each charge state (i.e. a difference of rates),
weighted by the occupancy of the state.  The direct contribution
to $\gammaup$ neglects any possible coherence between successive
excitation events; as a result, it fails to recover the classical
expression of Eq. (\ref{ClassicalSQ}) in the small-$\Omega$ limit.

We now consider the ``interference" contribution to $\gammaup$
coming from the second term in the expression for matrix $Q$ (c.f.
Eq. (\ref{QDefn})).
After some algebra, we obtain the following for the interference
contribution to $\gammaup$:
\begin{equation}
    \gammaup |_{int} =
    - \frac{2 A^2}{\Omega^2}
    \left( p_0 \Gamma^{-}_{10} +
                p_1 \Gamma^{-}_{01} \right)
    \frac{\left(\Gamma_{\Sigma}\right)^2}
        { \Omega^2 + \left( \Gamma_{\Sigma} \right)^2 }
                \label{TotalInt}
\end{equation}
where:
\begin{equation}
    \Gamma_{\Sigma} \equiv    \frac{\Gamma^{-}_{10} + \Gamma^{+}_{10} +
    \Gamma^{-}_{01} +
    \Gamma^{+}_{01}}{2}
\end{equation}
This contribution is purely negative, and is only significant
(relative to the direct contribution) at low frequencies $\Omega <
\Gamma$.
We can interpret this equation as describing the interference
between {\it two} consecutive excitation events. For example,
consider the first term in Eq. (\ref{TotalInt}). This describes a
process where a SET initially in the charge state $n=0$ undergoes
a tunnel event to the $n=1$ state, creating a superposition of
qubit ground and excited states.  At some later time the SET
relaxes to the stationary distribution $(p_0,p_1)$ of the charge
states, again partially exciting the qubit.  Letting $\Delta t$
represent the time between these two events, we have the
approximate sequence:
\begin{eqnarray}
    | 0,g_0 \rangle & \mapright{\Gamma_{10}^-} &
        | 1,g_1 \rangle + \alpha | 1,e_1
        \rangle  \nonumber \\
     & \mapright{\Delta t } &
        e^{i \Omega \Delta t / 2} | 1,g_1 \rangle +
        e^{-i \Omega \Delta t / 2} \alpha | 1,e_1 \rangle
         \label{IntState} \\
    & \mapright{\Gamma_{\Sigma}} &
        \left(e^{i \Omega \Delta t / 2} \beta -
        e^{-i \Omega \Delta t / 2} \alpha \right)
        | 0,e_0 \rangle + ... \label{FinalState}
\end{eqnarray}
Here, $\alpha$ is the amplitude associated with qubit excitation
having occurred during the first ($n=0 \ra 1$) tunnel event, while
$\beta$ is the amplitude associated with excitation occurring
during the second ($n=1 \ra 0$) tunneling.  These amplitudes will
be given by the corresponding matrix overlap elements:
\begin{equation}
    \alpha  =  \langle e_1 | g_0 \rangle \simeq \frac{A}{\Omega}
    \hspace{1 cm}
    \beta  =  \langle e_0 | g_1 \rangle \simeq -\frac{A}{\Omega}
    \label{EQOverlapElements}
\end{equation}
In the final state after the two tunnelings (Eq.
(\ref{FinalState})), there are two terms in the amplitude of the
state $| 0,e_0 \rangle$, corresponding to the fact that qubit
excitation could have occurred in either the first or the second
tunnel event. To get a rate for this double excitation event, we
should take the modulus squared of the final $| 0,e_0 \rangle$
state amplitude, then multiply by the occupancy of the initial
state ($p_0$) and the rate of the first tunnel event
($\Gamma_{10}$). The interference term in the resulting expression
takes the form:
\begin{equation}
    \gammaup|_{int} = (p_0 \Gamma_{10}) \times 2 \textrm{Re} \left(
        \alpha^* \beta e^{i \Omega \Delta t}  \right)
    = -(p_0 \Gamma_{10}) \frac{2 A^2}{\Omega^2} \cos(
    \Omega \Delta t )
    \label{InterferenceTerm}
\end{equation}
The above expression is a function of the time $\Delta t$ between
the first and second tunnel events.  This time is determined by
the fact that the intermediate superposition state (Eq.
(\ref{IntState})) corresponds to a non-stationary distribution of
charge on the SET island, and will decay via tunneling to the
stationary distribution $(p_0,p_1)$ at a rate $\Gamma_{\Sigma}$.
Taking this decay to be Poissonian, and averaging over $\Delta t$,
we obtain:
\begin{equation}
    \gammaup|_{int} = -(p_0 \Gamma_{10} ) \frac{2 A^2}{\Omega^2}
        \frac{  \left(\Gamma_{\Sigma} \right)^2 } { \Omega^2 +
        (\Gamma_{\Sigma})^2 }
\end{equation}
This is precisely the first term in Eq. (\ref{TotalInt}); the
second term can be obtained in the same way, by now considering a
situation where the SET is initially in the $n=1$ charge state. As
claimed, $\gammaup|_{int}$ corresponds to the interference between
two consecutive excitation events.  The negative sign of this
contribution can be directly traced to the matrix overlap elements
(c.f. Eq.~\ref{EQOverlapElements}).  Also, we see that the
suppression of the interference term at large $\Omega$ results
from phase randomization occurring during the delay time between
the two excitation events.

Returning to the total noise, we combine Eq. (\ref{TotalInt}) with
the direct contribution Eq. (\ref{Direct}) to $\gammaup$;
comparing against Eq. (\ref{FGR}), we obtain the final expression
for $S_Q(\Omega)$ at all {\it negative} frequencies:
\begin{equation}
    S_Q(-|\Omega|) =
        \frac{p_0 \Gamma^{-}_{10} + p_1 \Gamma^{-}_{01}}
            { \Omega^2 + \frac{1}{4}
                \left(
            \Gamma^{-}_{10} + \Gamma^{+}_{10} + \Gamma^{-}_{01} +
                \Gamma^{+}_{01} \right)^2  }
    \label{SQNegative}
\end{equation}
Note for large $|\Omega|$ (i.e. $|\Omega| > \max(\Delta
E^{\alpha}_{01},\Delta E^{\alpha}_{10}) \simeq V_{DS}/2)$,
$S_Q(-\Omega)$ will vanish identically at zero temperature.
Physically, this cutoff corresponds to the largest amount of
energy the SET can give up to the qubit during a single tunnel
event; giving up more energy would suppress the event completely
(i.e. the tunnel rates have a step-function form at zero
temperature, c.f. Eq. (\ref{gammadefn})). If one included higher
order processes in the tunneling (i.e. went beyond sequential
tunneling), correlated tunneling events involving the full voltage
drop over both junctions, $V_{DS}$, would move this cutoff to
higher values of absolute frequency.

We now turn to the relaxation rate $\gammadown$, and hence the
positive frequency parts of $S_Q$.  The calculation proceeds
exactly as that for $\gammaup$, the only modification being that
one now needs to include the charge states $n=2$ and $n=-1$, as
the SET could absorb enough energy from the qubit to make
transitions to these states possible.  We can combine the result
for $\gammadown$ with Eq. (\ref{SQNegative}) to obtain a single,
compact expression for the noise at all frequencies first obtained
by Johansson {\it et.~al} \cite{Wendin}: \footnote{Eq.
(\ref{FullSQ}) ignores additional order $g/(2 \pi)$ terms which
arise in the denominator at positive frequencies large enough to
turn on tunneling to higher charge states; such terms are clearly
negligible in the sequential tunneling regime due to the smallness
of $g/(2 \pi)$.}
\begin{equation}
    S_{Q}(\omega) = \frac
        {p_0
            \left[\Gamma_{10}(\omega) + \Gamma_{-1,0}(\omega)
            \right]  +
        p_1
            \left[\Gamma_{01}(\omega) + \Gamma_{21}(\omega)
            \right]
        }
        { \omega^2 + \frac{1}{4}
                \left[
            \Gamma_{10}(\omega) + \Gamma_{10}(-\omega)
            +\Gamma_{01}(\omega) + \Gamma_{01}(-\omega)
        \right]^2 }
    \label{FullSQ}
\end{equation}

\begin{figure}
\includegraphics{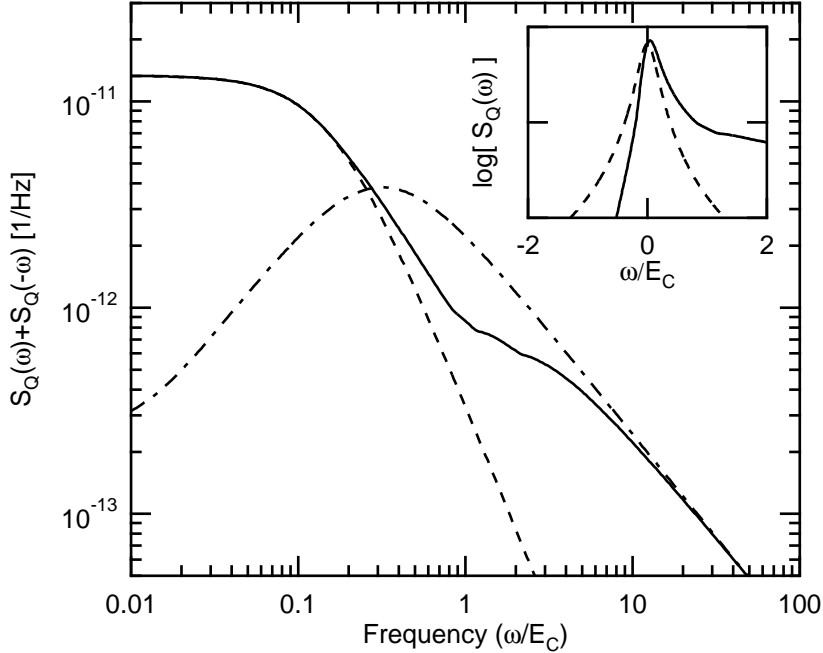}
\caption{Symmetrized SET charge noise as a function of frequency,
for typical SET parameters: $g=1$, $E_C / k_B = 2 {\textrm K}$,
$\NN = 0.33$, $e V_{DS} = E_C$, and $T = 20 {\textrm mK}$. The
dashed line is the classical telegraph noise
(Eq.~(\ref{ClassicalSQ})), while the dot-dashed line is the noise
of two parallel tunnel junctions (Eq. (\ref{Nyquist})). Inset:
full (non-symmetrized) quantum noise for identical SET parameters;
the dashed line is the symmetric classical telegraph noise.}
\label{SETChgNoiseFig}
\end{figure}

\begin{figure}
\includegraphics{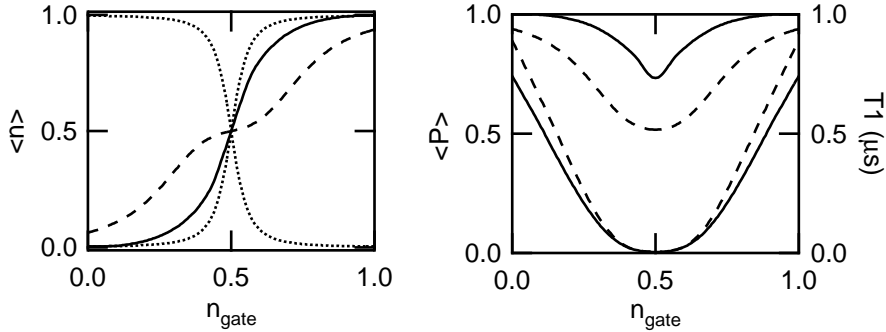}
\caption{ a) Average charge state of a Cooper pair box coupled to
a SET, as a function of box gate voltage, using identical SET
parameters as above.  The box parameters are $E_C / k_B = 0.5
{\textrm K}$, $E_J / k_B = 0.25 {\textrm K}$ and the coupling
constant is $\kappa = 0.04$.  We also include relaxation effects
due to a $10\%$ coupling to a $50 \Omega$ environment. The dashed
curve corresponds to assuming the SET produces classical telegraph
noise, the solid curve corresponds to using the full quantum noise
of the SET, and the dashed-dot curve is the box ground state. b)
The relaxation time $T_1$ for the same system, as a function of
box gate voltage.} \label{BoxPlusSETFigure}
\end{figure}

Shown in Figure~\ref{SETChgNoiseFig} is the symmetrized noise
$S_Q(\omega) + S_Q(-\omega)$ for typical SET parameters.  One can
clearly see abrupt changes in the slope of this curve; each of
these kinks corresponds to a threshold frequency at which a given
tunneling process either turns on or turns off.  For comparison,
curves corresponding to classical telegraph noise and to the
uncorrelated noise of two tunnel junctions are also shown. At low
frequencies the symmetrized true noise matches the classical
curve; for higher frequencies, it lies {\it above} the classical
curve but {\it below} the curve corresponding to the uncorrelated
case.  The inset of this figure shows the both the negative and
positive frequency parts of $S_Q(\omega)$.

It is easy to check that in the limit $\omega \ra 0$, Eq.
(\ref{FullSQ}) recovers the classical telegraph expression of Eq.
(\ref{ClassicalSQ}).  In the high-frequency, zero temperature
limit, one can also see that Eq.~(\ref{FullSQ}) approaches the
uncorrelated result of Eq.~(\ref{Nyquist}) from {\it below}:
\begin{equation}
S_Q(\omega) \ra \Theta(\omega) \frac
    { 4
        \left(\frac{g}{2 \pi}\right)
         \omega \left(1 -  \frac{E_C}{2 \hbar \omega} \right) }
    { \omega^2 + \frac{g^2}{\pi^2} \omega^2 }
    \rightarrow
    \Theta(\omega) \frac{2 g }{\pi \omega}
\end{equation}
Note that at high frequencies, it is only the ``direct" terms
which contribute to the noise-- the interference contribution is
not important in the limit of uncorrelated tunneling.  The fact
that the noise approaches the high frequency limit from below
results from the tendency of charging energy induced correlations
(which are present for a finite $\omega / E_C$) to suppress
fluctuations of $n$, and thus suppress the noise. Note that the
interpolation between the low and high frequency limits here is
very different than, e.g., interpolating between thermal noise and
zero-point fluctuation noise in a tunnel junction.  In the latter
case, one is effectively {\it combining} two sources of noise;
here, one is simply turning off correlations brought on by the
charging energy by increasing $\omega$.

Finally, shown in Figure~\ref{BoxPlusSETFigure} is the average
charge state of a Cooper pair box coupled to a SET with identical
parameters to that in Fig.~\ref{SETChgNoiseFig}.  We have also
included here the relaxation effects of the environment, modelled
as in Sec.~\ref{BoxOhmicSection} as a $50 \Omega$ impedance. Note
that even near the box degeneracy point, there are large
deviations between the result obtained using the full quantum
noise of the SET and that obtained from using only classical
telegraph noise. In Fig.~\ref{BoxPlusSETFigure}b, we show the
relaxation rate $T_1$ for the same system.  Note that the
differences between using the full quantum noise and the classical
expression are not so evident here.

\section{Summary}
In this article, we have emphasized the need to discuss quantum
noise processes using their two-sided spectral densities. Because
of the quantum nature of noise, the positive and negative
frequencies are generally unequal, in order to account for
spontaneous emission. A two-level system was shown to be an ideal
spectrum analyzer for probing the quantum nature of a noise
process or reservoir. With the advent of real electrical circuits
which behave as coherent two-level systems (e.g.,
\cite{Vion},\cite{Lehnert}), we can now build and use {\it quantum
electrical} spectrum analyzers. We also described the use of a
qubit as a {\it theoretical} tool, by following the evolution of
the density matrix of a TLS coupled to the noise-producing system
of interest. This technique appears to be quite powerful, as it
can yield analytical results for the full quantum noise spectrum
of a wide variety of devices, including the superconducting SET
\cite{Aash}. The distinction between the classical noise and the
quantum noise, found in this way, leads to dramatically different
predictions (c.f.~Fig.~\ref{BoxPlusSETFigure} and
Ref.~\cite{Aash}) for continuous measurements of qubits with an
SET. The ``coupled-system" calculational approach also allows
predictions of the dephasing by the measurement, the performance
relative to the Heisenberg uncertainty limit
\cite{NatureBackaction}, the fidelity of single-shot measurements
of the qubit states, and the effects of strong coupling to the
qubit. The combined theoretical and experimental advances raise
many interesting possibilities for testing our understanding of
quantum measurement theory with mesoscopic devices.

\begin{acknowledgements}
The authors acknowledge the generous support of this work by the
NSA and ARDA under ARO contracts ARO-43387-PH-QC (RS,SG) and
DAAD19-02-1-044 (MD), by the NSF under DMR-0196503 \& DMR-0084501
(AC,SG), the David and Lucile Packard Foundation (RS), and the
W.M. Keck Foundation.
\end{acknowledgements}

\end{document}